\documentclass[%
preprint,
superscriptaddress,
amsmath,amssymb,
pra,
]{revtex4-1}

\usepackage{graphicx}
\usepackage{dcolumn}
\usepackage{bm}
\usepackage{rotating}

\usepackage{amsthm}
\usepackage{amsmath}
\usepackage{amssymb}
\usepackage{graphicx}
\usepackage{url}
\usepackage{float}
\usepackage{bm}
\usepackage{ifthen}
\usepackage[usenames,dvipsnames]{color}
\usepackage{mathrsfs}
\usepackage[colorlinks=true,citecolor=blue,urlcolor=black]{hyperref}
\usepackage{float}
\usepackage{subfigure}
\usepackage{enumitem}
\usepackage{xcolor}
\usepackage{setspace}
\usepackage{braket}

\usepackage{lipsum}

\usepackage{verbatim}
\usepackage{array}
\usepackage{booktabs}
\usepackage{longtable}
\usepackage{multirow}

\newcommand{\be}{\begin{equation}}
	\newcommand{\ee}{\end{equation}}

\usepackage[mathlines]{lineno}

\begin{document}
	
	\newpage
	
	\title{Experimental demonstration of high-rate measurement-device-independent quantum key distribution over asymmetric channels}

	\author{Hui Liu}
	 \thanks{These authors contributed equally to this work.}
	\affiliation{Shanghai Branch, Hefei National Laboratory for Physical Sciences at Microscale and Department of Modern Physics, University of Science and Technology of China, Shanghai, 201315, China}
	\affiliation{CAS Center for Excellence and Synergetic Innovation Center in Quantum Information and Quantum Physics, University of Science and Technology of China, Shanghai 201315, P. R. China}
	
	\author{Wenyuan Wang}
	 \thanks{These authors contributed equally to this work.}
	\affiliation{Centre for Quantum Information and Quantum Control (CQIQC), Dept. of Electrical \& Computer Engineering and Dept. of Physics, University of Toronto, Toronto,  Ontario, M5S 3G4, Canada}
	
	\author{Kejin Wei}
	\affiliation{Shanghai Branch, Hefei National Laboratory for Physical Sciences at Microscale and Department of Modern Physics, University of Science and Technology of China, Shanghai, 201315, China}
	\affiliation{CAS Center for Excellence and Synergetic Innovation Center in Quantum Information and Quantum Physics, University of Science and Technology of China, Shanghai 201315, P. R. China}

	\author{Xiao-Tian Fang}
	\affiliation{Shanghai Branch, Hefei National Laboratory for Physical Sciences at Microscale and Department of Modern Physics, University of Science and Technology of China, Shanghai, 201315, China}
	\affiliation{CAS Center for Excellence and Synergetic Innovation Center in Quantum Information and Quantum Physics, University of Science and Technology of China, Shanghai 201315, P. R. China}

	\author{Li Li}
	\affiliation{Shanghai Branch, Hefei National Laboratory for Physical Sciences at Microscale and Department of Modern Physics, University of Science and Technology of China, Shanghai, 201315, China}
	\affiliation{CAS Center for Excellence and Synergetic Innovation Center in Quantum Information and Quantum Physics, University of Science and Technology of China, Shanghai 201315, P. R. China}

	\author{Nai-Le Liu}
	\affiliation{Shanghai Branch, Hefei National Laboratory for Physical Sciences at Microscale and Department of Modern Physics, University of Science and Technology of China, Shanghai, 201315, China}
	\affiliation{CAS Center for Excellence and Synergetic Innovation Center in Quantum Information and Quantum Physics, University of Science and Technology of China, Shanghai 201315, P. R. China}

	\author{Hao Liang}
	\affiliation{Shanghai Branch, Hefei National Laboratory for Physical Sciences at Microscale and Department of Modern Physics, University of Science and Technology of China, Shanghai, 201315, China}
	\affiliation{CAS Center for Excellence and Synergetic Innovation Center in Quantum Information and Quantum Physics, University of Science and Technology of China, Shanghai 201315, P. R. China}

	\author{Si-Jie Zhang}
	\affiliation{Shanghai Branch, Hefei National Laboratory for Physical Sciences at Microscale and Department of Modern Physics, University of Science and Technology of China, Shanghai, 201315, China}
	\affiliation{CAS Center for Excellence and Synergetic Innovation Center in Quantum Information and Quantum Physics, University of Science and Technology of China, Shanghai 201315, P. R. China}

	\author{Weijun Zhang}
	\affiliation{State Key Laboratory of Functional Materials for Informatics, Shanghai Institute of Microsystem and Information Technology, Chinese Academy of Sciences, Shanghai 200050, China}

	\author{Hao Li}
	\affiliation{State Key Laboratory of Functional Materials for Informatics, Shanghai Institute of Microsystem and Information Technology, Chinese Academy of Sciences, Shanghai 200050, China}

	\author{Lixing You}
	\affiliation{State Key Laboratory of Functional Materials for Informatics, Shanghai Institute of Microsystem and Information Technology, Chinese Academy of Sciences, Shanghai 200050, China}

	\author{Zhen Wang}
	\affiliation{State Key Laboratory of Functional Materials for Informatics, Shanghai Institute of Microsystem and Information Technology, Chinese Academy of Sciences, Shanghai 200050, China}

	\author{Hoi-Kwong Lo}
	\affiliation{Centre for Quantum Information and Quantum Control (CQIQC), Dept. of Electrical \& Computer Engineering and Dept. of Physics, University of Toronto, Toronto,  Ontario, M5S 3G4, Canada}
		
	\author{Teng-Yun Chen}
	\affiliation{Shanghai Branch, Hefei National Laboratory for Physical Sciences at Microscale and Department of Modern Physics, University of Science and Technology of China, Shanghai, 201315, China}
	\affiliation{CAS Center for Excellence and Synergetic Innovation Center in Quantum Information and Quantum Physics, University of Science and Technology of China, Shanghai 201315, P. R. China}
	
	\author{Feihu Xu}
	\affiliation{Shanghai Branch, Hefei National Laboratory for Physical Sciences at Microscale and Department of Modern Physics, University of Science and Technology of China, Shanghai, 201315, China}
	\affiliation{CAS Center for Excellence and Synergetic Innovation Center in Quantum Information and Quantum Physics, University of Science and Technology of China, Shanghai 201315, P. R. China}	
		
    \author{Jian-Wei Pan}
	\affiliation{Shanghai Branch, Hefei National Laboratory for Physical Sciences at Microscale and Department of Modern Physics, University of Science and Technology of China, Shanghai, 201315, China}
	\affiliation{CAS Center for Excellence and Synergetic Innovation Center in Quantum Information and Quantum Physics, University of Science and Technology of China, Shanghai 201315, P. R. China}

\begin{abstract}
Measurement-device-independent quantum key distribution (MDI-QKD) can eliminate all detector side channels and it is practical with current technology. Previous implementations of MDI-QKD all use two symmetric channels with similar losses. However, the secret key rate is severely limited when different channels have different losses. Here we report the results of the first high-rate MDI-QKD experiment over \emph{asymmetric} channels. By using the recent 7-intensity optimization approach, we demonstrate $>$10x higher key rate than previous best-known protocols for MDI-QKD in the situation of large channel asymmetry, and extend the secure transmission distance by more than 20-50 km in standard telecom fiber. The results have moved MDI-QKD towards widespread applications in practical network settings, where the channel losses are asymmetric and user nodes could be dynamically added or deleted.
\end{abstract}
	
\maketitle

Quantum key distribution (QKD) promises information-theoretical security in communications~\cite{bb84, e91}. In practice, however, the realistic QKD implementations might introduce device imperfections~\cite{xu2019}, which deviate from the idealized models~\cite{Zhaoyi,Feihu2010,Makarov2010,Makarov2011,Weinfurter2011,Leuchs}. Among many protocols to resolve the device imperfections~\cite{mayers1998quantum,acin2007device,braunstein2012side}, the measurement-device-independent QKD (MDI-QKD) protocol~\cite{Lomdi} has attracted a lot of research interests due to its practicality with current technology and its nature advantage of immunity to all detector attacks. Experimental MDI-QKD~\cite{Rub2013,Silva2013,Liu2013,Tang2014,yanlin2014,Wang2015,Zhiyuan2016} has advanced significantly up to a distance of 404 km in low loss fiber~\cite{Yin2016} and a key rate of 1 Mbits/s~\cite{Comandar}. Many theoretical improvements have been proposed to guarantee the practical security~\cite{Ma2012,wang2013three,Feihu2013,Curty2014,mdiparameter,mdifourintensity}. Notably, the recent proposal of twin-field QKD has the capability to overcome the rate-distance limit of QKD~\cite{lucamarini2018overcoming}.

\begin{figure}[t!]
	\includegraphics[scale=0.46]{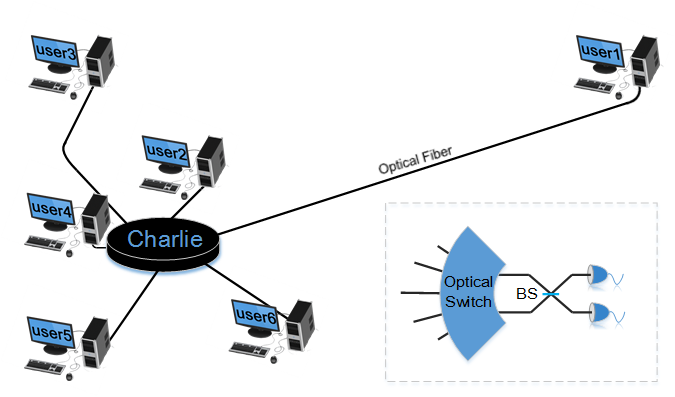}
	\caption{An illustration of a star-type MDI-QKD network providing six users with access to the untrusted relay, Charlie. Inset: an example of the possible implementation by Charlie.}
	\label{fig:1}
\end{figure}

The future of QKD is believed to be a quantum network in which many user nodes are connected together via quantum channels and centric servers, such as the star-type network illustrated in Fig.~\ref{fig:1}. MDI-QKD is well suited to construct such a centric QKD network even with an untrusted relay, i.e., the six users in Fig.~\ref{fig:1} can securely communicate with each other, though Charlie is insecure. Such a MDI-QKD network, as demonstrated in~\cite{Yanlin2016}, presents a huge advantage over traditional trusted-relay based QKD networks~\cite{quantumnetwork1,quantumnetworkchen,quantumnetwork2}.


In a practical quantum network, it is inevitable that some users are further away from the central relay, while the others are closer to the relay. For instance, in Fig.~\ref{fig:1}, user 1 and user 3 are farther from Charlie than the other users. This topology has appeared naturally in previous field QKD networks~\cite{quantumnetwork1,quantumnetworkchen,quantumnetwork2}. Unfortunately, so far, all implementations to MDI-QKD have been performed either through near-symmetric channels~\cite{Silva2013,Liu2013,Tang2014,yanlin2014,Wang2015,Zhiyuan2016} or through the deliberate addition of loss in one channel to balance the total losses in the two arms~\cite{Rub2013}. However, the assumption of near-symmetric channels is clearly an unsatisfactory situation in a practical MDI-QKD network. Adding loss in one channel will severely limit the key rate and the secure distance in the asymmetric setting~\cite{Feihu2013}, and it also means that the addition/deletion of a new node will inevitably affect every other existing node, which is highly inconvenient and limits the scalability of the network.


We for the first time demonstrate high-rate MDI-QKD over asymmetric channels and achieve substantially higher key rate over previous methods for MDI-QKD. Our experiment employs the 7-intensity optimization method proposed recently in~\cite{placeholder}. We demonstrate that the 7-intensity method can be implemented in software only without having to physically modify any channel, and it is highly scalable that can be easily integrated into existing quantum network infrastructure.



In asymmetric MDI-QKD with two users, Alice and Bob have channel transmittances $\eta_{A}$ and $\eta_{B}$ ($\eta_{A}\neq\eta_{B}$). The main question is how to choose the optimal intensities of the weak coherent pulses for Alice and Bob, denoted by $s_A$ and $s_B$, so as to maximize the key rate~\cite{Feihu2013}. A natural option is to choose the intensities to balance the channel losses, i.e., $s_A\eta_{A}=s_B\eta_{B}$. By doing so, a symmetry of photon flux can arrive at Charlie, and thus resulting in a good Hong-Ou-Mandel (HOM) dip~\cite{HOM1987}. The dependency of HOM visibility versus balance of photon number flux can be seen in~\cite{moschandreou2018experimental,placeholder}. However, such an option is sub-optimal, and may even generate no key rate at all when the channel asymmetry is high. The fundamental reason is that MDI-QKD is related but \emph{different} from HOM dip. That is, HOM dip affects only the errors in $X$ basis (i.e., the phase error rate estimated with decoy state method), but has no effect to errors in $Z$ basis (i.e., the bit error rate). Therefore, the optimal method is to decouple the decoy state estimation in $X$ basis from key generation in $Z$ basis. This is the key idea of the 7-intensity optimization method proposed in~\cite{placeholder}. Note that Ref.~\cite{mdifourintensity} also mentioned on passing the possibility of using different intensities for Alice and Bob, but no analysis on this important asymmetric case was performed there.

In the 7-intensity optimization method~\cite{placeholder}, Alice and Bob each selects a set of 4 intensities, namely signal state \{$s_A$, $s_B$\} in the Z basis, and decoy states \{$\mu_A, \nu_A, \omega$\} and \{$\mu_B, \nu_B, \omega$\} in the $X$ basis, respectively. The parameters that Alice and Bob choose include 7 different intensities in total, as well as the proportions to send them. The secret key is generated only from the $Z$ basis, while the data in the $X$ basis are all used to perform the decoy state analysis. The decoy state intensities are chosen to compensate for asymmetry and ensure good HOM visibility in the $X$ basis (and roughly satisfy ${\mu_A \over \mu_B}={\nu_A \over \nu_B} \approx {\eta_B \over \eta_A}$, which maintains symmetry of photon flux arriving at Charlie). On the other hand, the signal state is decoupled from the decoy states, and can be freely adjusted to maximize key rate in the Z basis (and generally ${s_A \over s_B} \neq {\eta_B \over \eta_A}$). Overall, Alice and Bob optimize 12 implementation parameters: $[s_A,\mu_A,\nu_A,p_{s_A},p_{\mu_A},p_{\mu_A},s_B,\mu_B,\nu_B,p_{s_B},p_{\mu_B},p_{\mu_B}].$ To efficiently choose the optimal parameters, we use a local search algorithm and follow the optimization technique in~\cite{placeholder}, which converts the 12 parameters into polar coordinate and searches them while locking the decoy state intensities at: ${\mu_A \over \nu_A}={\mu_B \over \nu_B}$~\cite{supp}. The optimization technique is highly efficient, and takes less than 0.1s for each run of full optimization on a common desktop PC (with a quad-core Intel i7-4790k processor, using parallelization with 8 threads).


\begin{figure*}[ht!]
	\includegraphics[scale=0.5]{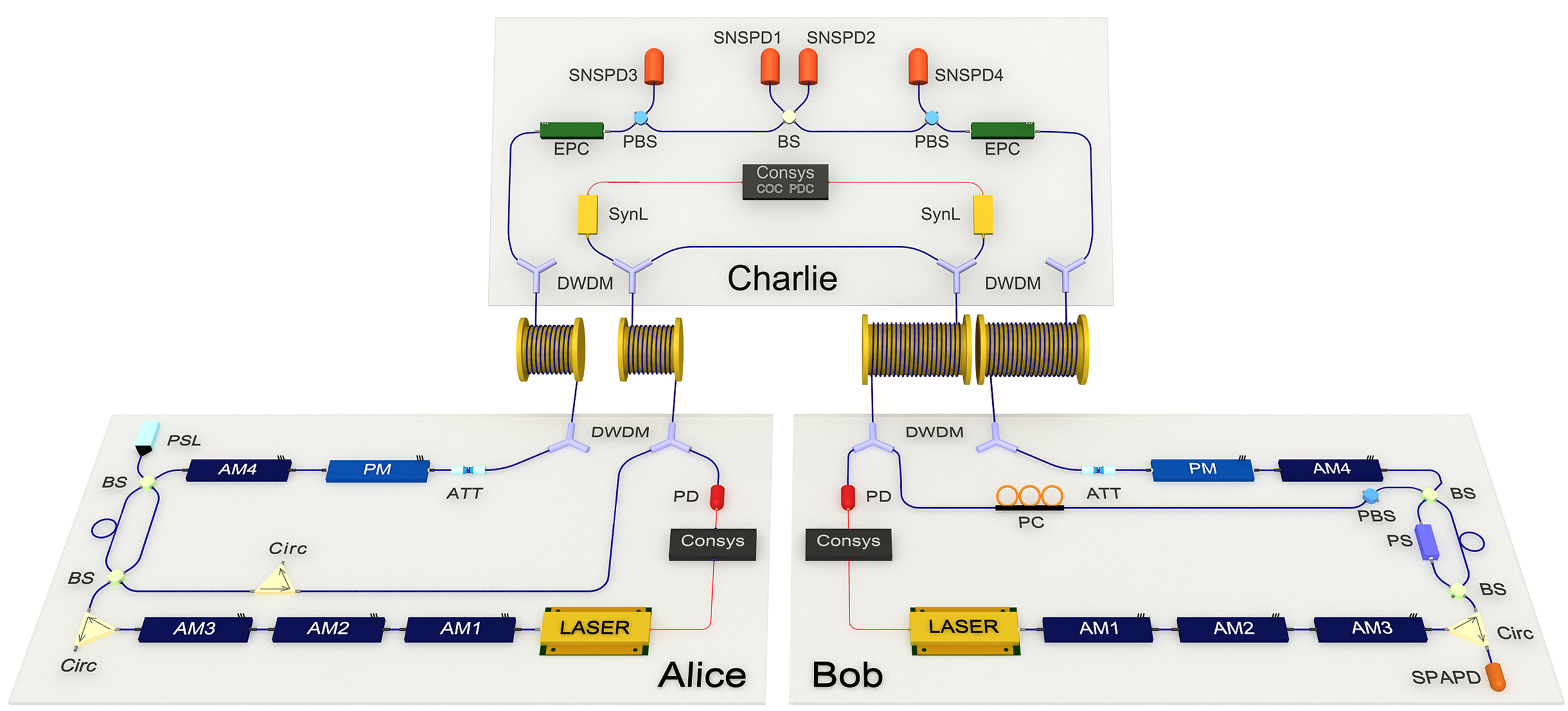}
	\caption{MDI-QKD setup. Alice's (Bob's) signal laser pulses are modulated into signal and decoy intensities by three amplitude modulators (AM1-AM3). Key bits are encoded by a Mach-Zehnder interferometer, AM4, and a phase modulator (PM). In Charlie, the polarization stabilization system in each link includes an electric polarization controller (EPC), a polarization beam splitter (PBS) and a superconducting nanowire single-photon detector (SNSPD); the Bell state measurement (BSM) system includes a 50/50 beam splitter (BS), SNSPD1 and SNSPD2. Abbreviations of other components: DWDM, dense wavelength division multiplexer; ConSys, control system; ATT, attenuator; PSL, phase-stabilization laser; Circ, circulator; PC, polarization controller; PS, phase shifter; SPAPD, single-photon avalanche photodiode.}
	\label{fig:setup}
\end{figure*}

\begin{table}[ht]
\caption{List of parameters characterized from experiment: detector dark count rate $Y_0$, detector system efficiency $\eta_d$, optical misalignment $e_d^X$, $e_d^Z$ in the X and Z bases, fiber loss coefficient $\alpha$ in dB/km, error-correction efficiency $f$, security parameter $\epsilon$, and the total number of laser pulses $N$ sent by Alice/Bob. }

\begin{tabular}{p{2cm}<{\centering}p{0.7cm}<{\centering}p{0.7cm}<{\centering}p{0.7cm}<{\centering}p{0.7cm}<{\centering}p{0.7cm}<{\centering}p{0.8cm}<{\centering}p{0.8cm}<{\centering}}
\hline\hline

\specialrule{0em}{1.5pt}{1.5pt}
   $Y_0$ &     $\eta_d$ &      $e_d^Z$ &     $e_d^X$ &    $\alpha$ &     $f$ &    $\epsilon$ &     $N$ \\
\specialrule{0em}{1.5pt}{1.5pt}
\hline
\specialrule{0em}{1.2pt}{1.2pt}
  $6.40\times10^{-8}$ &       46\% &     0.5\% &        4\% &       0.19 &       1.16 &   $10^{-10}$ &   $10^{12}$ \\
\specialrule{0em}{1.2pt}{1.2pt}
\hline\hline
\end{tabular} \label{Tab1}
\end{table}

To implement MDI-QKD over two asymmetric channels, we construct a time-bin-phase encoding MDI-QKD setup in Fig.~\ref{fig:setup}. Alice and Bob each possesses an internally modulated laser which emits phase-randomized laser pulses at a clock rate of 75 MHz. The gain-switched laser diode can naturally generate optical pulses with random phases. AM1 (amplitude modulator) is used to tailor the pulse shape by cutting off the overshoot rising edge of laser pulses. AM2 and AM3 are employed to randomly modulate the intensities of signal state and weak decoy states. The time-bin encoding is implemented by utilizing a combination of a Mach-Zehnder interferometer (MZI), AM4 and a phase modulator (PM). For $Z$ basis, the key bit is encoded in time bin $\ket{0}$ or $\ket{1}$ by AM4, while for the $X$ basis, it is encoded in the relative phase 0 or $\pi$ by the PM. Alice and Bob send their laser pulses through two standard fiber spools, $L_{\text{A}}$ and $L_{\text{B}}$, to Charlie, who performs Bell state measurement (BSM). The BSM includes a 50/50 beam splitter (BS) and two superconducting nanowire single-photon detectors (SNSPD1 and SNSPD2). The main system parameters characterized in the experiment are shown in Table~\ref{Tab1}.

To compensate for the relative phase drift and establish a common phase reference, Alice employs a phase-stabilization laser (PSL) and Bob employs a phase shifter (PS) in one of the arms of his MZI and a single-photon avalanche photodiode. To properly interfere the two pulses at Charlie, we develop a real-time polarization feedback control system, an automatic time calibration system and a temperature feedback control system~\cite{supp}. Thanks to the feedback control systems, the observed visibility of the two photon interference is about $46\%$ and the system has a long-term stability over tens of hours. This stability enables us to collect a large number of signal detections, thus properly considering the finite-key effect~\cite{Curty2014}.

\begin{figure*}[ht]
	\includegraphics[scale=0.46]{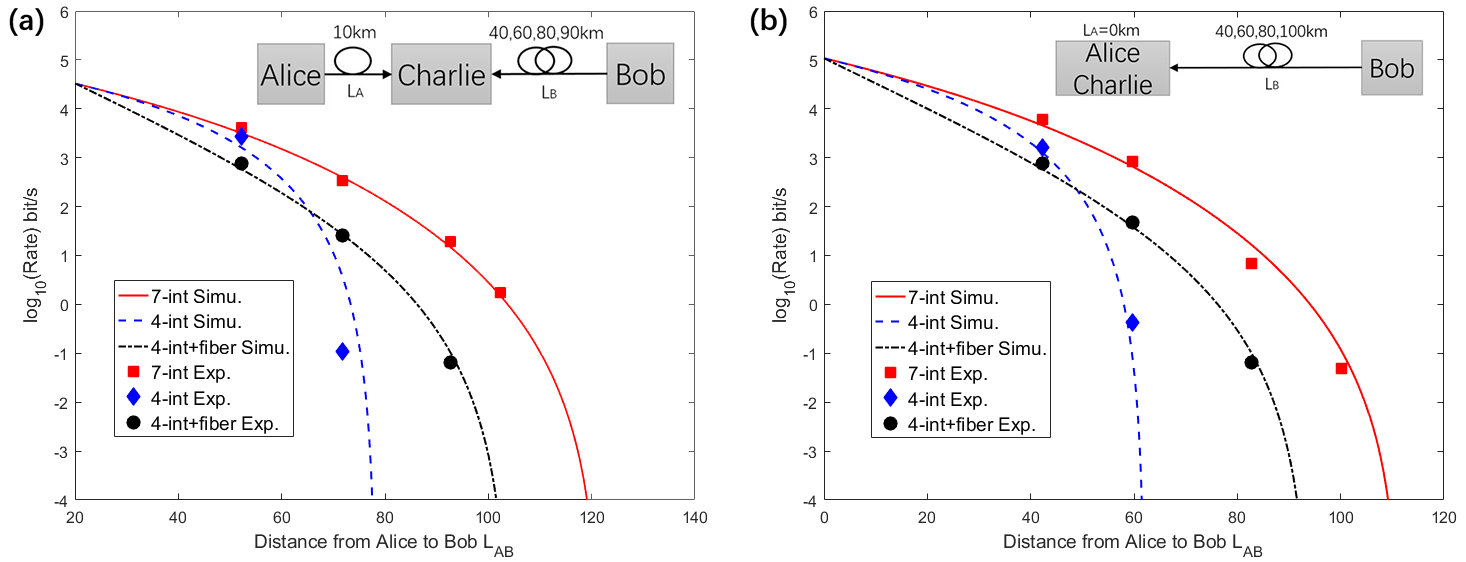}
	\caption{Simulation (curve) and experiment results (data points) for secret rate (bit/pulse) vs the total distance $L_{\text{AB}}$ in standard telecom fiber. \textbf{(a)} $L_{\text{A}}$ is fixed at 10 km, while $L_{\text{B}}$ is selected at 40, 60, 80, 90 km. (b) $L_{\text{A}}$ is fixed at 0 km, while $L_{\text{B}}$ is selected at 40, 60, 80, 100 km. The points (curves) in the figure indicate the experimental (simulation) results for (i) 4-intensity method shown in blue diamond points (blue dashed line), where the same intensities and proportions for Alice and Bob are selected and optimized in the 4-intensity protocol~\cite{mdifourintensity,Yin2016}; (ii) 4-intensity+fiber method~\cite{Rub2013} shown in black circle points (black dot-dash line);(iii) 7-intensity method~\cite{placeholder}, shown in red square points (red solid line). As can be seen, for the 4-intensity methods, adding fibers improves the key rate in long distances, but it does not in short distances. In contrast, the 7-intensity method always achieves substantially higher key rate than any of the other two methods, especially when channel asymmetry is high.
}
	\label{fig:results}
\end{figure*}

\begin{table}[h]
\centering
\caption{Example implementation parameters and experimental results for $L_{\text{A}}$=10 km and $L_{\text{B}}$=60 km. $s_{11}^Z$ is the estimated yield of single photons in the $Z$ basis and $e_{11}^X$ is the estimated phase-flip error rate of single photons in the $X$ basis. $Q_{ss}^Z$ and $E_{ss}^Z$ are the observed gain and QBER for signal states. $R$ is the secret key rate (bit/s). Ratio is the key rate advantage of the 7-intensity method over the given method.}
\begin{tabular}{cccc}
\hline
\hline
\specialrule{0em}{1.5pt}{1.5pt}
Parameters & 7-intensity & 4-intensity & 4-intensity+fiber \\
\specialrule{0em}{1.5pt}{1.5pt}
\hline
       $s_A$ &     0.169  &    0.119  &    0.363  \\
       $s_B$ &     0.614  &     0.119  &     0.363  \\
       $\mu_A$ &    0.056  &    0.180  &    0.280  \\
       $\mu_B$ &     0.465  &     0.180  &     0.280  \\
       $\nu_A$ &     0.011  &     0.023  &     0.058 \\
       $\nu_B$ &     0.089  &     0.023  &     0.058  \\
    $p_{s_A}$ &     0.599  &     0.256  &     0.483  \\
    $p_{s_B}$ &     0.600  &     0.256  &     0.483  \\
    $p_{\mu_A}$ &     0.030  &     0.035  &     0.045  \\
    $p_{\mu_B}$ &     0.031  &     0.035  &     0.045  \\
    $p_{\nu_A}$ &     0.254  &     0.490  &     0.320  \\
    $p_{\nu_B}$ &     0.248  &     0.490  &     0.320  \\
\specialrule{0em}{1.5pt}{1.5pt}
     $s_{11}^Z$ &   $1.63\times 10^{-3}$ &   $1.97\times 10^{-3}$ &   $1.86\times 10^{-4}$ \\
\specialrule{0em}{1.5pt}{1.5pt}
     $e_{11}^X$ &    14.00\% &    20.28\% &    16.72\% \\
\specialrule{0em}{1.5pt}{1.5pt}
  $Q_{ss}^Z$ &   $2.24\times 10^{-4}$ &   $3.05\times 10^{-5}$ &   $3.10\times 10^{-5}$ \\
\specialrule{0em}{1.5pt}{1.5pt}
  $E_{ss}^Z$ &     0.91\% &     2.50\% &     0.91\% \\
\specialrule{0em}{1.5pt}{1.5pt}
         R &   $343$ &   $0.11$ &   $25.50$ \\
         Ratio & 1 & 3118 & 13.5 \\
\hline
\hline
\end{tabular} \label{Tab2}
\end{table}

We implement the 7-intensity method over different choices of channel lengths~\cite{supp}. First, we fix the distance between Alice and Charlie at 10 km, i.e., $L_{\text{A}}=10$ km, while the distance between Bob and Charlie $L_{\text{B}}$ varies from 40 km to 90 km. At each channel setting, we use the system parameters listed in Table~\ref{Tab1} to perform a numerical optimization on the implementation parameters, based on three optimization strategies: (i) 4-intensity method, where the same intensities and proportions for Alice and Bob are selected and optimized in the 4-intensity protocol~\cite{mdifourintensity,Yin2016}; (ii) 4-intensity+fiber method, where the asymmetry of channels is first compensated by adding additional losses~\cite{Rub2013} and then the same intensities and proportions for Alice and Bob are selected; (iii) 7-intensity method. The results are shown in Fig.~\ref{fig:results}(a). 7-intensity method can substantially increase the key rate and maximum distance of MDI-QKD in the case of high channel asymmetry: at $L_{\text{B}}=60$ km, the 7-intensity method generates a secret key rate of over an order of magnitude higher than the 4-intensity+fiber method, and extends the maximum distance for approximately 20km compared to 4-intensity+fiber method, and 40km compared to 4-intensity method alone.

Next, we demonstrate for the first time a ``single-arm" MDI-QKD, as shown in the inset figure in Fig.~\ref{fig:results}(b), where we place Alice and Charlie at the same location, i.e., $L_{\text{A}}=0$ km. $L_{\text{B}}$ varies from 40 km to 100 km. The results are shown in Fig.~\ref{fig:results}(b). Such a single-arm setup only uses one public channel, and could be highly useful in free-space QKD, where Alice and Bob typically have a single free-space channel, in the middle of which adding a relay is unfeasible (e.g., ship-to-ship or satellite-ground channel). In this case, adding fiber in the lab would also be inconvenient due to turbulence or moving platforms. Using ``single-arm" MDI-QKD, however, Bob can place a relay in his lab, such that Alice and Bob can enjoy the security of MDI-QKD through this channel, and maintain satisfactory key rate.

We list the implementation parameters and the main experimental results for $L_{\text{A}}=10$ km and $L_{\text{B}}=60$ km in Table~\ref{Tab2}. Note that the parameters in 7-intensity method are quite different from those two types of 4-intensity methods. We obtain a secret key rate of 343 bits/s with 7-intensity method, which is 13.5 times higher than that of 4-intensity+fiber method. By using the joint-bound analysis~\cite{mdifourintensity}, the key rate can be further improved to 645 bits/s. Moreover, the 7-intensity optimization method can greatly extend the transmission distance by about 50 km fiber. Furthermore, we also tested an extreme case where $L_A$=0km and $L_B$=100km. 7-intensity produces a secret key rate of 0.049 bit/s. In contrast, no key bits can be extracted with either strategy of using 4-intensity method with/without fiber.

The method of asymmetric intensities and decoupled bases we demonstrated can be applied to general quantum information protocols. First, the asymmetric method is important to the future implementation of free-space MDI-QKD with a moving relay such as satellite. For instance, the channel transmittances in satellite-based quantum communication are constantly changing with up to 20-dB channel mismatch~\cite{yin2017satellite}. Second, the asymmetric method can be readily applied to MDI quantum digital signature (QDS)~\cite{roberts2017experimental,yin2017experimental} and twin-field (TF) QKD~\cite{lucamarini2018overcoming}. The key generation formula of MDI-QDS is similar to that of MDI-QKD, where the proposed method can be directly implemented~\cite{supp}. TF-QKD relies on single-photon interference, where the intensity-asymmetry affects both the interference visibility and the single-photon gain~\cite{supp}. Our methods of asymmetric choice of intensities and optimization of parameters can be implemented to improve the key rate for asymmetric TF-QKD~\cite{zhou2019asymmetric}. However, we note that the two encoding bases are symmetric in TF-QKD, thus the method of decoupled bases might not be applicable~\cite{supp}. Finally, other protocols that rely on single-photon or two-photon interference, such as comparison of coherent states~\cite{andersson2006experimentally} and quantum fingerprinting~\cite{arrazola2014quantum,xu2015experimental,guan2016observation}, can also be benefited from our methods when they are working in an asymmetric setting.

In conclusion, by using the recent 7-intensity method, we demonstrate an order of magnitude higher key rate and an extension of 20-50 km distance over previous best-known MDI-QKD protocols. While previous methods of adding fibers inconveniently require the modification of every existing node with the addition/deletion of a new node, our 7-intensity method implements the optimization in software only and provides much better scalability. Overall, our results have moved MDI-QKD towards a more practical network setting, where the channel losses can be asymmetric and nodes can be dynamically added or deleted.


\begin{acknowledgements}
This work has been supported by the National Key R\&D
Program of China, National Natural Science Foundation
of China, Anhui Initiative in Quantum Information
Technologies, Fundamental Research Funds for the Central
Universities, Shanghai Science and Technology Development Funds. F. X.
acknowledges the support from Thousand Young Talent
program of China. W. W. and H.-K. L. were supported by
NSERC, U.S. Office of Naval Research, CFI, ORF, and
Huawei Canada.
\end{acknowledgements}

\newpage

\appendix
\section{Optimization Algorithm}

Here we briefly describe the optimization algorithm, as proposed in Ref.~\cite{placeholder}, for the intensities and the probabilities of sending them in the 7-intensity optimization method.

As described in the main text, there are a total of 12 parameters that need to be optimized:

\begin{equation}
[s_A,\mu_A,\nu_A,p_{s_A},p_{\mu_A},p_{\mu_A},s_B,\mu_B,\nu_B,p_{s_B},p_{\mu_B},p_{\mu_B}]
\end{equation}

To navigate such a large parameter space, a local search algorithm is necessary. However, the key rate function versus these parameters is in fact non-smooth, hence local search does not work well in this case. To address this, we need to convert the parameters to polar coordinate:

\begin{equation}
[r_s,\theta_s, r_\mu, r_\nu, \theta_{\mu\nu}, p_{s_A},p_{\mu_A},p_{\mu_A},p_{s_B},p_{\mu_B},p_{\mu_B}]
\end{equation}

\noindent where the conversion follows that:

\begin{equation}
	\begin{aligned}
	r_{s}&=\sqrt{{s_A}^2+{s_B}^2}\\
	\theta_s&=arctan({s_A\over s_B})
	\end{aligned}
\end{equation}

\noindent and the same applies for $r_\mu,\theta_{\mu\nu}$ and $r_\nu,\theta_{\mu\nu}$. Note that here we have locked the polar angle of $\mu$ and $\nu$ to be always equal. This is because, the optimal values of the decoy intensities for Alice and Bob always satisfy~\cite{placeholder}

\begin{equation}
{\mu_A \over \nu_A}={\mu_B \over \nu_B}
\end{equation}

\noindent for the 7-intensity optimization method. The intensity probabilities are not involved in non-smoothness and therefore do not need to be changed. With the 11 parameters now, we can perform a local search algorithm, such as coordinate descent \cite{placeholder,mdiparameter}, to efficiently find the set of optimal intensities. Coordinate descent algorithm alternatively optimizes each variable at a time while keeping others constant. When all variables are searched, the algorithm starts over again with the first variable. With enough iterations, this algorithm can reach a local maximum point. For asymmetric MDI-QKD, the algorithm can find the global maximum point (which is the only local maximum).

\section{The Detail of MDI-QKD Systems}

Alice and Bob each possess an internally modulated laser which emits phase-randomized laser pulses at a clock rate of 75 MHz. The pulse width is about 3.4 ns, and the center wavelength is at 1550.12 nm with a full width at half maximum (FWHM) of about 15 pm. AM4, is used to modulate the vacuum state, where the extinction ratio between the signal state and the vacuum state is larger than 23 dB. The MZI divides each incoming pulse into two time-bins with 6.4 ns time interval. The superconducting nanowire single-photon detectors (SNSPD1 and SNSPD2) have detection efficiency~70\% and dark count rate ~30 cps. Due to an extra insertion loss~1.2 dB in Charlie and 15\% non-overlap between laser pulse and detection time window, the total system detection efficiency is~46\%.

To get a high visibility of two-photon interference , we adopt several feedback systems~\cite{yanlin2014} to calibrate the polarization, time and spectrum modes of the signal pulses generated by two independent laser sources, as shown in Fig. 2 in main text.

First, to make sure that the polarization of two pulses are indistinguishable, we plug a EPC and a PBS before the BS in the BSM. Intensities of the reflection port of the PBS are monitored by a SNSPD, which outputs a feedback signal to control the EPC to minimize the intensities.

Besides, to precisely overlap timing mode of the two signal pulses, there are two calibration processes in the experiments. First, Charlie generates two  synchronization lasers(Synls, 1570nm) pulses which are synchronized  by a crystal oscillator circuit (COC). The pulses are respectively sent to Alice and Bob, who detected it by a PD. The output signals of the PDs are used to synchronize lasers in Alice and Bob. Second, Alice and  Bob respectively send their signal laser pulses to Charlie, who use the SNSPD to measure the arriving time of the pulses. Then, by using a programmable delay chip, Charlie adjusts the time delay between the two SynL according to the arriving time difference. The total timing calibration precision is below 10ps and the arriving time jitter of SNSPD is below 50ps, both them are far smaller that the pulse width of 2.4 ns.

For the spectrum mode, we first use an optical spectrum analyzer (OSA,YOKOGAWA AQ6370B) to calibrate Alice's signal pulses and phase feedback pulses wavelength at 1550.12nm. Then we observe the Hong-Ou-Mandel (HOM) interference of two pulses in Charlie and adjust the operating temperature of the laser in Bob to the value where the HOM dip are found.

Alice and Bob need to have a shared phase reference frame which fluctuates with temperature and stress. Using a PSL, Alice sends laser pulses from her AMZI to Bob's AMZI via an additional link between Alice and Bob. Bob monitors the power at one of outputs of his interferometer with a SPAPD and then minimizes the counts of SPAPD by using a PS in Bob's AMZI.

\section{Application to asymmetric quantum digital signature}

In MDI quantum digital signature (QDS)~\cite{2016puthoor,roberts2017experimental,yin2017experimental}, there exists a sufficiently large signature length which makes the protocol secure, if the condition
\begin{equation}\label{1}
Q_{Z}^{0,0}+Q_{Z}^{1,1}[1-h(e_{X}^{1,1})]-h(E_Z)>0
\end{equation}
holds. Here, $Q_{Z}^{i,i}$ is the lower bound on the count rate when Alice and Bob sent $Z$-basis pulses containing $i$ photons, $e_X^{1,1}$ is the upper bound for the single-photon phase error rate and $E_Z$ is the quantum bit error rate (QBER) in $Z$ basis.

Apparently, Eq.~\eqref{1} is the same as the key rate formula of MDI-QKD~\cite{Lomdi} except for that MDI-QDS omits the inefficient factor of error correction. Hence, we can directly apply the proposed asymmetric method to improve the performance of MDI-QDS over asymmetric channels.

\section{Application to asymmetric twin-field QKD}

Unlike MDI-QKD, in the case of TF-QKD \cite{lucamarini2018overcoming}, the key bits are generated from singles $\ket{01}+\ket{10}$ rather than coincidences $\ket{11}$. Alice and Bob send signals whose phases are announced and post-selected to be in the same "phase-slice", and the signals are received by a third party, Eve, who announces the detector events at detectors $C$ and $D$. The key rate can be written as \cite{lucamarini2018overcoming},
\begin{equation}
	R={d\over M}\{Q_1[1-h_2(e_1)]-fQ_\mu h_2(E_\mu)\}
\end{equation}

\noindent where $M$ is the number of phase slices, $d$ is a phase slice post-selection factor, $Q_1,e_1$ are the estimated gain and phase-error rate of single photons, $Q_\mu,E_\mu$ are the observed gain and QBER for the signal states, and $f$ is the error correction efficiency.

Conceptually, for single photons, the asymmetry between transmittances in the two channels decreases the visibility of single-photon interference, resulting in higher QBER. In reality, Alice and Bob both use WCP sources. Consider Alice and Bob sending intensities $\mu_A$ and $\mu_B$. Let us define

\begin{equation}
\begin{aligned}
\gamma_A&=\sqrt{\mu_A\eta_A\eta_d}\\
\gamma_B&=\sqrt{\mu_B\eta_B\eta_d}
\end{aligned}
\end{equation}

\noindent where $\eta_A,\eta_B$ are the channel transmittances between Alice (Bob) and Charlie, and $\eta_d$ is the detector efficiency. Following \cite{Feihu2013}, the received intensities $D_C,D_D$ at the detectors (here for simplicity we ignore the dark counts) are:
\begin{equation}
	\begin{aligned}
	D_C&=(\gamma_A^2+\gamma_B^2-2\gamma_A\gamma_Bcos\phi)/2\\
	D_D&=(\gamma_A^2+\gamma_B^2+2\gamma_A\gamma_Bcos\phi)/2\\
	\end{aligned}
\end{equation}

\noindent where $\phi$ is the relative phase between Alice's and Bob's signals. Consider the detector $C$, the visibility can be written as:
\begin{equation}
\begin{aligned}
	v_1={{I_{max}-I_{min}}\over{I_{max}+I_{min}}}={{2\gamma_A\gamma_B}\over{\gamma_A^2+\gamma_B^2}}={{2}\over{{k}+{1\over k}}} \\
\end{aligned}
\end{equation}
\noindent where $k={\sqrt{{\mu_A\eta_A}\over{\mu_B\eta_B}}}$ is the ratio between arriving intensities at Eve's beam splitter. $v_1$ is a function that reaches maximum $1$ when $k=1$, and monotonically decreases with $k$ when $k$ deviates from $1$. In fact, we can compare this with the two-photon interference visibility from two WCP sources as used in MDI-QKD:
\begin{equation}
\begin{aligned}
v_2=1-{{P_{coin}}\over {P_C}{P_D}}={{2\mu_A\eta_A\mu_B\eta_B}\over{(\mu_A\eta_A+\mu_B\eta_B)^2}}={{2}\over{2+k^2+{1 \over k^2}}} \\
\end{aligned}
\end{equation}
\noindent where $P_C,P_D,P_{coin}$ are respectively the count probability of detector $C$, detector $D$, and the coincidence count probability.

\begin{figure}[h]
\includegraphics[scale=0.25]{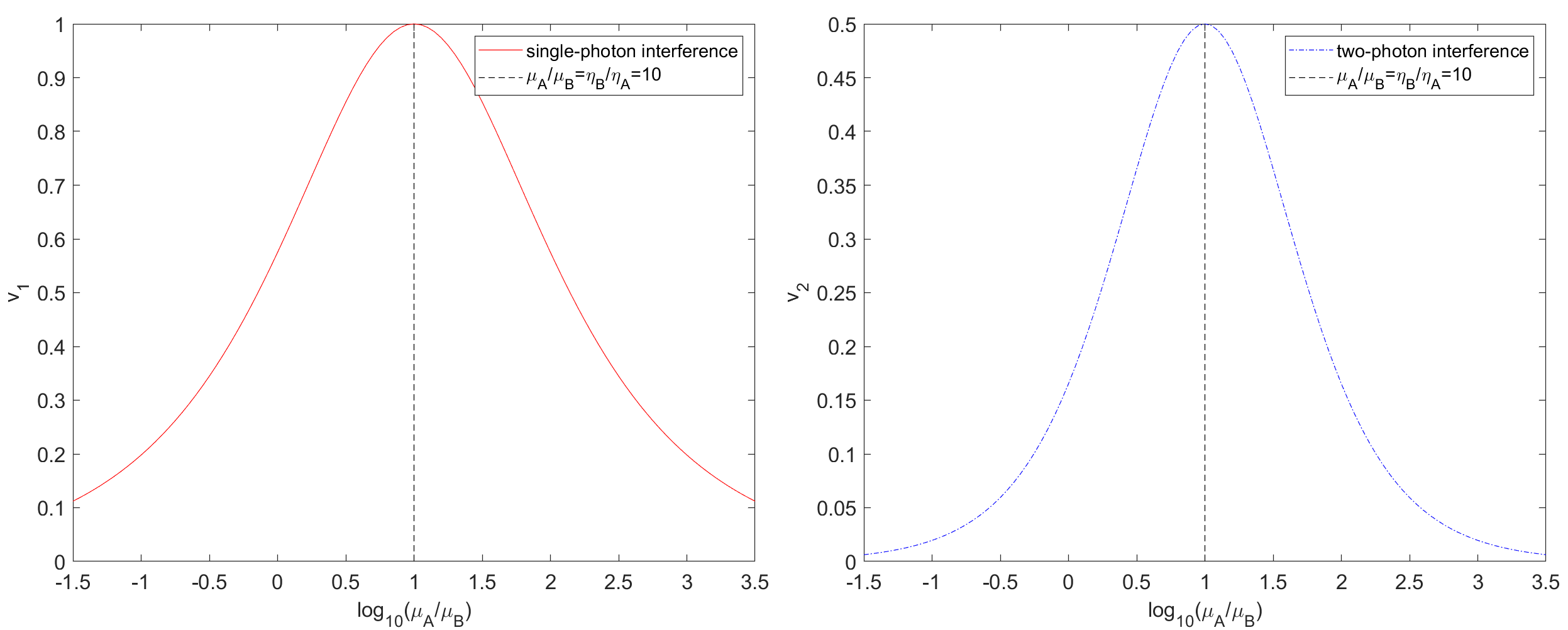}
\caption{The visibility for a single-photon interference (used in TF-QKD) and for a two-photon interference (used in MDI-QKD) versus the ratio of intensities used by Alice and Bob, using WCP sources. Here we consider two channels with 50km difference in standard fiber (i.e. with $\eta_A/\eta_B=0.1$). The visibility is the highest in both cases when Alice and Bob use intensities that satisfy $\mu_A\eta_A=\mu_B\eta_B$, and the visibility drops as ratio of arriving intensities at Charles becomes imbalanced. Note that a two-photon interference with WCP sources can only reach a maximum of 50\% visibility while single-photon interference can reach 100\%. Therefore, just like for MDI-QKD, the visibility of single-photon interference in TF-QKD (and as a result its QBER) heavily depends on the balance of intensities.}
\label{fig:visibility}
\end{figure}

We have plotted both visibilities versus ratio of arriving intensities in Fig.~\ref{fig:visibility}. We can see that just like for MDI-QKD, the single-photon interference visibility in TF-QKD heavily depends on the balance of arriving intensities. The visibility of the interference between the two pulses directly affects the observed QBER for both the signal and decoy states. Therefore, we can also apply our method to compensate for channel asymmetry (such as using $\eta_A\mu_A=\eta_B\mu_B$) and obtain higher single-photon interference visibility and lower QBER.

Note that the single-photon gain $Q_1$ also depends on the intensity values~\cite{zhou2019asymmetric}. Hence an asymmetric choice of intensities, i.e., $\eta_A\mu_A=\eta_B\mu_B$, may not be the optimal method. A careful optimization of the parameters, following the approach in \cite{placeholder}, are likely required to produce the optimal key rate in asymmetric TF-QKD. Note also that the decoupling of basis might not work in TF-QKD. The reason is that, unlike MDI-QKD where the two bases are inherently asymmetric (Charles only measures in $Z$ basis~\cite{placeholder}), for TF-QKD the two bases are symmetric, and both will depend on the visibility of interference. Some extensions of the initial TF-QKD protocol use only one basis $X$ for encoding and another basis $Z$ for decoy-state analysis, where decoupling of bases and using different intensity choices potentially might work, but a rigorous study of this will be the subject of future studies.

\bibliography{bi}


\onecolumngrid
\begin{center}
\begin{sidewaystable}
\centering
\caption{List of main parameters and experimental results. Here, $s_{11}^Z$ is the yield of single photon states in the $Z$ basis and $e_{11}^X$ is  the phase-flip error rate of single photon states. $Q_{s_As_B}^Z$ and $E_{s_As_B}^Z$ are the observed yield and QBER for source $s_As_B$, respectively. R is the final key rate. }
\begin{tabular}{c|ccc|ccc|cc|c}
\hline
\hline
\multicolumn{1}{c|}{\multirow{2}{2cm}{Parameters $L_A$=10km}} &         \multicolumn{3}{c|}{$L_B$=42km} &         \multicolumn{3}{c|}{$L_B$=62km} & \multicolumn{2}{c|}{$L_B$=83km} &      \multicolumn{1}{c}{$L_B$=92km} \\
\cline{2-10}
\multicolumn{1}{c|}{} & 7-int & 4-int & 4-int+fiber & 7-int & 4-int & 4-int+fiber & 7-int & 4-int+fiber & 7-int \\
\hline
       $s_A$ &       0.330  &     0.390  &     0.454  &     0.169  &     0.119  &     0.363  &     0.059  &     0.209  &     0.027  \\
       $s_B$ &       0.626  &     0.390  &     0.454  &     0.614  &     0.119  &     0.363  &     0.544  &     0.209  &     0.462  \\
       $\mu_A$ &     0.101  &     0.210  &     0.252  &     0.056  &     0.180  &     0.280  &     0.031  &     0.316  &     0.023  \\
       $\mu_B$ &     0.383  &     0.210  &     0.252  &     0.465  &     0.180  &     0.280  &     0.542  &     0.316  &     0.580  \\
       $\nu_A$ &     0.019  &     0.032  &     0.050  &     0.011  &     0.023  &     0.058  &     0.006  &     0.068  &     0.004  \\
       $\nu_B$ &     0.071  &     0.032  &     0.050  &     0.089  &     0.023  &     0.058  &     0.105  &     0.068  &     0.114  \\
     $p_{s_A}$ &     0.720  &     0.648  &     0.657  &     0.599  &     0.256  &     0.483  &     0.423  &     0.220  &     0.306  \\
     $p_{s_B}$ &     0.721  &     0.648  &     0.657  &     0.600  &     0.256  &     0.483  &     0.423  &     0.220  &     0.306  \\
    $p_{\mu_A}$ &    0.020  &     0.020  &     0.027  &     0.030  &     0.035  &     0.045  &     0.046  &     0.074  &     0.057  \\
    $p_{\mu_B}$ &    0.020  &     0.020  &     0.027  &     0.031  &     0.035  &     0.045  &     0.048  &     0.074  &     0.060  \\
    $p_{\nu_A}$ &    0.176  &     0.228  &     0.214  &     0.254  &     0.490  &     0.320  &     0.369  &     0.480  &     0.446  \\
    $p_{\nu_B}$ &    0.176  &     0.228  &     0.214  &     0.248  &     0.490  &     0.320  &     0.352  &     0.480  &     0.419  \\

     $s_{11}^Z$ &   4.26E-03 &   4.62E-03 &   1.17E-03 &   1.63E-03 &   1.97E-03 &   1.86E-04 &   7.40E-04 &   2.73E-05 &   4.33E-04 \\
     $e_{11}^X$ &    8.69\% &    10.35\% &    12.58\% &    14.00\% &    20.28\% &    16.72\% &    17.90\% &    24.26\% &    20.82\% \\
  $Q_{s_As_B}^Z$ &  1.02E-03 &   7.38E-04 &   2.60E-04 &   2.24E-04 &   3.05E-05 &   3.10E-05 &   2.85E-05 &   1.78E-06 &   7.17E-06 \\
  $E_{s_As_B}^Z$ &   0.89\% &     1.19\% &     0.81\% &     0.91\% &     2.50\% &     0.91\% &     1.03\% &     0.80\% &     0.89\% \\
         $R$ &      5.57E-05 &   3.69E-05 &   1.02E-05 &   4.57E-06 &   1.45E-09 &   3.40E-07 &   2.59E-07 &   8.70E-10 &   2.30E-08 \\
\hline
\multicolumn{1}{c|}{$L_A$=0km} &         \multicolumn{3}{c|}{$L_B$=42km} &         \multicolumn{3}{c|}{$L_B$=60km} & \multicolumn{2}{c|}{$L_B$=83km} &      \multicolumn{1}{c}{$L_B$=100km} \\
\cline{2-10}
\hline
       $s_A$ &       0.299  &     0.352  &     0.454  &     0.143  &     0.040  &     0.373  &     0.047  &     0.209  &     0.008  \\
       $s_B$ &       0.663  &     0.352  &     0.454  &     0.638  &     0.040  &     0.373  &     0.567  &     0.209  &     0.360  \\
       $\mu_A$ &     0.074  &     0.197  &     0.252  &     0.040  &     0.160  &     0.277  &     0.022  &     0.316  &     0.012  \\
       $\mu_B$ &     0.407  &     0.197  &     0.252  &     0.476  &     0.160  &     0.277  &     0.545  &     0.316  &     0.612  \\
       $\nu_A$ &     0.014  &     0.026  &     0.050  &     0.007  &     0.017  &     0.057  &     0.004  &     0.068  &     0.002  \\
       $\nu_B$ &     0.074  &     0.026  &     0.050  &     0.089  &     0.017  &     0.057  &     0.105  &     0.068  &     0.120  \\
     $p_{s_A}$ &     0.735  &     0.622  &     0.657  &     0.613  &     0.101  &     0.500  &     0.443  &     0.220  &     0.202  \\
     $p_{s_B}$ &     0.735  &     0.622  &     0.657  &     0.614  &     0.101  &     0.500  &     0.443  &     0.220  &     0.203  \\
    $p_{\mu_A}$ &    0.018  &     0.018  &     0.027  &     0.028  &     0.035  &     0.043  &     0.044  &     0.074  &     0.068  \\
    $p_{\mu_B}$ &    0.019  &     0.018  &     0.027  &     0.029  &     0.035  &     0.043  &     0.046  &     0.074  &     0.070  \\
    $p_{\nu_A}$ &    0.167  &     0.248  &     0.214  &     0.246  &     0.599  &     0.310  &     0.357  &     0.480  &     0.517  \\
    $p_{\nu_B}$ &    0.167  &     0.248  &     0.214  &     0.240  &     0.599  &     0.310  &     0.340  &     0.480  &     0.477  \\
     $s_{11}^Z$ &    6.84E-03 &   5.19E-03 &   1.17E-03 &   2.82E-03 &   3.34E-03 &   2.22E-04 &   9.58E-04 &   2.73E-05 &   4.17E-04 \\
     $e_{11}^X$ &    10.23\% &     9.78\% &    12.58\% &    11.25\% &    15.39\% &    14.70\% &    20.02\% &    24.26\% &    25.02\% \\
  $Q_{s_As_B}^Z$ &   1.31E-03 &   8.38E-04 &   2.60E-04 &   2.98E-04 &   6.32E-06 &   4.03E-05 &   3.49E-05 &   1.78E-06 &   1.76E-06 \\
  $E_{s_As_B}^Z$ &   1.00\% &     1.58\% &     0.81\% &     1.00\% &     2.80\% &     0.86\% &     1.03\% &     0.80\% &     0.80\% \\
         $R$ &     8.01E-05 &   2.21E-05 &   1.02E-05 &   1.14E-05 &   5.56E-09 &   6.29E-07 &   9.17E-08 &   8.70E-10 &   6.57E-10 \\
\hline
\hline
\end{tabular}
\end{sidewaystable}
\end{center}

\newpage
\begin{sidewaystable}[ht!]
\caption{List of the total gains and error gains of bell state $\psi^-$ in the cases of $L_A=10km$. The first row indicates the channel distance length and attenuation from Bob to Charlie. The notation $\alpha\beta$ shown in the second column denotes the pulse pair from Alice source $\alpha$ and Bob source $\beta$, respectively.}
\centering
\begin{tabular}{p{2cm}<{\centering} c|ccc|ccc|cc|c}
\hline
\hline
\multicolumn{2}{p{2cm}<{\centering}|}{$L_B$(Att)} &   \multicolumn{ 3}{c|}{42km(8.0dB)} &  \multicolumn{ 3}{c|}{62km(11.7dB)} & \multicolumn{ 2}{c|}{83km(15.7dB)} & 92km(17.5dB) \\
\hline
\multicolumn{ 2}{p{2cm}<{\centering}|}{Method} & 7-int & 4-int & 4-int+fiber & 7-int & 4-int & 4-int+fiber & 7-int & 4-int+fiber & 7-int \\
\hline
\multicolumn{1}{c}{\multirow{10}{0.5cm}{Total\\Gains}} &         $ss$ &  529982000 &  309687261 &  112210026 &   80413590 &    1997479 &    7242636 &    5100750 &      86140 &     671128 \\
\multicolumn{ 1}{c}{} &         $\mu\mu$ &     160464 &     271373 &     114406 &     105015 &     521423 &      66546 &      66076 &      43756 &      53137 \\
\multicolumn{ 1}{c}{} &         $\nu\nu$ &     448079 &     862716 &     304028 &     253796 &    1677973 &     155932 &     151134 &      85340 &     112771 \\
\multicolumn{ 1}{c}{} &         $\mu\nu$ &     518568 &    2411155 &     355724 &     340104 &    6263336 &     184777 &     217025 &     110651 &     166407 \\
\multicolumn{ 1}{c}{} &         $\nu\mu$ &     435644 &     331860 &     345149 &     262563 &     316203 &     181574 &     164018 &     101974 &     117519 \\
\multicolumn{ 1}{c}{} &         $\mu o$ &     177080 &    1031430 &     126082 &     129099 &    2765092 &      60339 &      82862 &      36891 &      69163 \\
\multicolumn{ 1}{c}{} &         $o\mu$ &     145394 &      62044 &     114283 &      79636 &      28252 &      61069 &      46348 &      31749 &      30971 \\
\multicolumn{ 1}{c}{} &         $\nu o$ &      68390 &     258891 &      41556 &      41171 &     616085 &      18131 &      25039 &      10464 &      21071 \\
\multicolumn{ 1}{c}{} &         $o\nu$ &      45319 &      14193 &      32111 &      22903 &       6997 &      19496 &      12268 &       9968 &       7985 \\
\multicolumn{ 1}{c}{} &         $oo$ &         34 &         21 &          5 &         18 &         22 &          7 &          8 &          2 &          3 \\
\hline
\multicolumn{1}{c}{\multirow{10}{0.5cm}{Error\\Gains}} &         $ss$ &    4704842 &    3691050 &     903743 &     734688 &      49958 &      65642 &      52352 &        692 &       5958 \\
\multicolumn{ 1}{c}{} &         $\mu\mu$ &      42156 &     101958 &      29481 &      29279 &     225881 &      17982 &      18427 &      12060 &      14618 \\
\multicolumn{ 1}{c}{} &         $\nu\nu$ &     123123 &     311694 &      83836 &      71032 &     717763 &      43523 &      42199 &      23685 &      31662 \\
\multicolumn{ 1}{c}{} &         $\mu\nu$ &     198094 &    1141819 &     135702 &     136799 &    3087498 &      68614 &      84966 &      41521 &      67831 \\
\multicolumn{ 1}{c}{} &         $\nu\mu$ &     157601 &      96533 &     127178 &      92640 &      88813 &      65551 &      58199 &      38153 &      40424 \\
\multicolumn{ 1}{c}{} &         $\mu o$ &      89600 &     520691 &      63013 &      65216 &    1419123 &      30128 &      41573 &      18375 &      34774 \\
\multicolumn{ 1}{c}{} &         $o\mu$ &      75143 &      29171 &      59294 &      39763 &      14031 &      29982 &      23279 &      15981 &      15308 \\
\multicolumn{ 1}{c}{} &         $\nu o$ &      35275 &     132892 &      20918 &      20934 &     309259 &       9277 &      12513 &       5208 &      10452 \\
\multicolumn{ 1}{c}{} &         $o\nu$ &      22520 &       7229 &      16029 &      11373 &       3387 &       9768 &       6126 &       5043 &       3991 \\
\multicolumn{ 1}{c}{} &         $oo$ &         30 &         12 &          2 &          2 &          6 &          4 &          0 &          0 &          1 \\
\hline
\hline
\end{tabular}
\end{sidewaystable}

\begin{sidewaystable}[ht!]
\caption{List of the total gains and error gains of bell state $\psi^-$ in the cases of $L_A=0km$. The first row indicates the channel distance length and attenuation from Bob to Charlie. The notation $\alpha\beta$ shown in the second column denotes the pulse pair from Alice source $\alpha$ and Bob source $\beta$, respectively.}
\centering
\begin{tabular}{p{2cm}<{\centering} c|ccc|ccc|cc|c}
\hline
\hline
\multicolumn{2}{p{2cm}<{\centering}|}{$L_B$(Att)} &   \multicolumn{3}{c|}{42km(8.0dB)} &  \multicolumn{ 3}{c|}{60km(11.4dB)} & \multicolumn{ 2}{c|}{83km(15.7dB)} & 100km(19.0dB) \\
\hline
\multicolumn{ 2}{p{2cm}<{\centering}|}{Method} & 7-int & 4-int & 4-int+fiber & 7-int & 4-int & 4-int+fiber & 7-int & 4-int+fiber & 7-int \\
\hline
\multicolumn{ 1}{c}{\multirow{10}{0.5cm}{Total\\Gains}} &         $ss$ &  709219832 &  324284028 &  112210026 &  112026597 &      64469 &   10064196 &    6843352 &      86140 &      72083 \\
\multicolumn{1}{c}{} &         $\mu\mu$ &     156842 &     352882 &     114406 &     111941 &     824061 &      74658 &      63990 &      43756 &      45564 \\
\multicolumn{1}{c}{} &         $\nu\nu$ &     452775 &    1308280 &     304028 &     283850 &    3198393 &     174730 &     142616 &      85340 &      87631 \\
\multicolumn{1}{c}{} &         $\mu\nu$ &     600030 &    4352714 &     355724 &     343732 &   12609257 &     197235 &     216230 &     110651 &     143354 \\
\multicolumn{1}{c}{} &         $\nu\mu$ &     449197 &     314469 &     345149 &     309693 &     431319 &     209512 &     141295 &     101974 &      90385 \\
\multicolumn{1}{c}{} &         $\mu\omega$ &     218789 &    1921910 &     126082 &     135845 &    5495396 &      64089 &      89197 &      36891 &      59855 \\
\multicolumn{1}{c}{} &         $\omega\mu$ &     149128 &      43637 &     114283 &      96747 &      28908 &      67637 &      37272 &      31749 &      22047 \\
\multicolumn{1}{c}{} &         $\nu\omega$ &      70166 &     469174 &      41556 &      44285 &    1233433 &      21570 &      25953 &      10464 &      17302 \\
\multicolumn{1}{c}{} &         $\omega\nu$ &      44174 &       9652 &      32111 &      27726 &       7571 &      20977 &      10636 &       9968 &       5469 \\
\multicolumn{1}{c}{} &         $\omega\omega$ &          7 &         29 &          5 &         22 &          7 &          9 &          7 &          2 &          3 \\
\hline
\multicolumn{1}{c}{\multirow{10}{0.5cm}{Error\\Gains}} &         $ss$ &    7113724 &    5119502 &     903743 &    1116248 &       1804 &      86126 &      70246 &        692 &        574 \\
\multicolumn{1}{c}{} &         $\mu\mu$ &      45820 &     143887 &      29481 &      29546 &     372745 &      19577 &      18010 &      12060 &      12710 \\
\multicolumn{1}{c}{} &         $\nu\nu$ &     125142 &     529786 &      83836 &      77686 &    1408517 &      47280 &      40233 &      23685 &      24750 \\
\multicolumn{1}{c}{} &         $\mu\nu$ &     248974 &    2086348 &     135702 &     137036 &    6229729 &      72300 &      89746 &      41521 &      59175 \\
\multicolumn{1}{c}{} &         $\nu\mu$ &     163484 &      81345 &     127178 &     113382 &     119372 &      76387 &      48685 &      38153 &      29803 \\
\multicolumn{1}{c}{} &         $\mu\omega$ &     109404 &     934790 &      63013 &      68947 &    2757313 &      32924 &      45632 &      18375 &      30019 \\
\multicolumn{1}{c}{} &         $\omega\mu$ &      73634 &      22163 &      59294 &      49024 &      14173 &      34585 &      18128 &      15981 &      10990 \\
\multicolumn{1}{c}{} &         $\nu\omega$ &      35015 &     236280 &      20918 &      22201 &     615180 &      10796 &      12780 &       5208 &       8669 \\
\multicolumn{1}{c}{} &         $\omega\nu$ &      21884 &       4937 &      16029 &      13939 &       3786 &      10430 &       5211 &       5043 &       2761 \\
\multicolumn{1}{c}{} &         $\omega\omega$ &          3 &         17 &          2 &         14 &          5 &          7 &          3 &          0 &          1 \\
\hline
\hline
\end{tabular}
\end{sidewaystable}

\end{document}